\documentclass[pre,twocolumn,aps,showpacs,nofootinbib,superscriptaddress]{revtex4-1} 

\usepackage{tikz}

\usepackage{amsmath}
\usepackage{graphicx}
\usepackage{dcolumn}
\usepackage{bm}
\usepackage{color}
\usepackage[normalem]{ulem}

\usepackage{amsfonts,epsfig,amsmath,amsthm,amssymb,graphics,verbatim}

\usepackage{color}

\newcommand{\be}{\begin{equation}}
\newcommand{\ee}{\end{equation}}
\newcommand{\bea}{\begin{eqnarray}}
\newcommand{\eea}{\end{eqnarray}}
\newcommand{\beann}{\begin{eqnarray*}}
\newcommand{\eeann}{\end{eqnarray*}}
\newcommand{\benn}{\begin{equation*}}
\newcommand{\eenn}{\end{equation*}}

\begin{document}

\title{An adaptive voter model on simplicial complexes}

\author{Leonhard Horstmeyer}
\affiliation{Section for the Science of Complex Systems, CeMSIIS, Medical University of Vienna, 
Spitalgasse 23, A-1090, Vienna, Austria}
\affiliation{Complexity Science Hub Vienna, Josefst{\"a}dterstrasse 39, A-1090 Vienna, Austria}
\author{Christian Kuehn}
\affiliation{Faculty of Mathematics, Technical University of Munich, Boltzmannstr.~3, 85748 Garching M\"unchen, Germany}
\affiliation{Complexity Science Hub Vienna, Josefst{\"a}dterstrasse 39, A-1090 Vienna, Austria}

\newtheorem{thm}{Theorem}[section]
\newtheorem{cor}[thm]{Corollary}
\newtheorem{prop}[thm]{Proposition}
\newtheorem{lem}[thm]{Lemma}
\newtheorem{ex}[thm]{Example}
\newtheorem{fact}[thm]{Fact}
\newtheorem{cjt}[thm]{Conjecture}
\newtheorem{rmrk}[thm]{Remark}

\newtheorem{defn}[thm]{Definition}

\newcommand{\cA}{{\mathcal A}}  
\newcommand{\cB}{{\mathcal B}}  
\newcommand{\cC}{{\mathcal C}}  
\newcommand{\cD}{{\mathcal D}}  
\newcommand{\cE}{{\mathcal E}}  
\newcommand{\cF}{{\mathcal F}}  
\newcommand{\cG}{{\mathcal G}}  
\newcommand{\cH}{{\mathcal H}}  
\newcommand{\cJ}{{\mathcal J}}  
\newcommand{\cK}{{\mathcal K}}  
\newcommand{\cL}{{\mathcal L}}  
\newcommand{\cM}{{\mathcal M}}  
\newcommand{\cN}{{\mathcal N}}  
\newcommand{\cO}{{\mathcal O}}  
\newcommand{\cP}{{\mathcal P}}  
\newcommand{\cQ}{{\mathcal Q}}  
\newcommand{\cR}{{\mathcal R}}  
\newcommand{\cS}{{\mathcal S}}  
\newcommand{\cT}{{\mathcal T}}  
\newcommand{\cU}{{\mathcal U}}  
\newcommand{\cV}{{\mathcal V}}  
\newcommand{\cW}{{\mathcal W}}  
\newcommand{\cX}{{\mathcal X}}  
\newcommand{\cY}{{\mathcal Y}}  
\newcommand{\cZ}{{\mathcal Z}}  

\input{motifs.tex}

\date{Version \today}

\begin{abstract}
Collective decision making processes lie at the heart of many social, 
political and economic challenges. The classical voter model is a well-established
conceptual model to study such processes. In this
work, we define a new form of adaptive (or co-evolutionary) voter model posed on a
simplicial complex, i.e., on a certain class of hypernetworks/hypergraphs. We
use the persuasion rule along edges of the classical voter model and the recently 
studied re-wiring rule of edges towards like-minded nodes, and introduce a new peer pressure
rule applied to three nodes connected via a 2-simplex. This simplicial adaptive voter
model is studied via numerical simulation. We show that adding the effect of peer
pressure to an adaptive voter model leaves its fragmentation transition, i.e., the
transition upon varying the re-wiring rate from a single majority state into to a 
fragmented state of two different opinion subgraphs, intact. Yet, above and below 
the fragmentation transition, we observe that the peer pressure has substantial 
quantitative effects. It accelerates the transition to a single-opinion state
below the transition and also speeds up the system dynamics towards fragmentation above the
transition. Furthermore, we quantify that there is a multiscale hierarchy in the model
leading to the depletion of 2-simplices, before the depletion of active edges. This 
leads to the conjecture that many other dynamic network models on simplicial complexes
may show a similar behaviour with respect to the sequential evolution of simplicies
of different dimensions.    
\end{abstract}


\maketitle

\section{Introduction}
\label{sec:introduction}

Contact and voter processes are a key theme of many disciplines including economics, 
epidemiology, mathematics, physics, and social science~\cite{Liggett}. One natural 
setting for these processes is an underlying (complex) network, or graph, on which a population 
of individuals interacts~\cite{BarratBarthelemyVespignani}. For example, in the context of 
opinion formation, which we focus upon here, two individuals/nodes/vertices may hold different 
opinions. For simplicity, let us assume that there are only two opinions possible, e.g., we can 
only vote for two possible parties. The classical voter model~\cite{HolleyLiggett,SoodRedner,
Dornicetal} describes the evolution of opinion dynamics on a fixed network by a 
Markov chain, either in discrete or continuous time; here we use the discrete-time variant. At each
time step, an edge is selected at random. If both vertices hold the same opinion, nothing happens. 
If they hold opposite opinions, then one adapts the opinion of the other with equal probability. Studying 
the long-time behaviour of such a dynamical system is already highly non-trivial as it
does depend crucially on the network structure; see e.g.~\cite{CastellanoViloneVespignani,
Gleeson1,NardiniKozmaBarrat,SoodAntalRedner,SoodRedner,SucheckiEguiluzSanMiguel,
VazquezEguiluz}.\medskip 

However, an important element of realism is missing in the classical voter model.
Social interactions in large populations almost never take place on a fixed network.
In fact, with whom we are in contact may also depend upon the difference or similarity
of opinions~\cite{McPhersonSmithLovinCook}. This viewpoint has led to the development of 
adaptive, or co-evolutionary, network models, in which there is interacting dynamics on and 
of the network; see e.g.~\cite{BornholdtRohlf,GrossDLimaBlasius,GrossSayama,JainKrishna2,
PachecoTraulsenNowak,KuehnZschalerGross,HorstmeyerKuehnThurner}. More precisely, 
the most common version of the adaptive voter model allows in addition to persuasion events 
also for re-wiring events, where edges between opposite-minded vertices are re-wired to edges 
between like-minded vertices. This makes the process much more realistic as it allows one to 
study via a relatively simple model complex self-adapting network structures. There is already 
a substantial literature on adaptive voter models~\cite{VazquezEguiluzMiguel,Bencziketal,BoehmeGross,Durrettetal,
Klamseretal,KozmaBarrat,RogersGross,Zschaleretal}.\medskip

Yet, recent changes in communication and social network formation cast serious doubt 
on the assumption that only binary interactions should matter in opinion 
formation~\cite{Jiaetal,WattsDodds}. These higher-order interactions have 
recently started to appear as a new focus in the analysis of complex network 
data sets~\cite{Bensonetal}. Yet, there are currently no available standard 
adaptive/co-evolutionary network dynamics models taking into account higher-order 
interactions. In this work, we propose and study a 
minimalistic extension of the adaptive voter model to include higher-order interaction 
between individuals. This model takes into account the well-known effect of peer-pressure, which
has been studied widely in many scientific fields~\cite{KandelLazear,BrownClasenEicher,HansenGraham}.
For example, if three individuals are connected in a friendship, and there is a 
disagreement in opinions, it is very likely that the majority opinion within the 
group of three prevails, i.e., a peer-pressure effect has occurred. To model whether 
a fully connected subgraph of three vertices is in a close-enough friendship or not, we 
need an additional structure beyond vertices and edges. A very general underlying structure 
would be to consider a hypergraph~\cite{Bretto} instead of a usual graph/network. Although this is possible, 
we are looking to develop a minimal and mathematically elegant formulation to capture the
essential effects of peer-pressure opinion dynamics. One natural choice in this context 
is to restrict ourselves to simplicial complexes~\cite{Hatcher}, where a triangle of connected nodes is in close friendship interaction if there is a 2-simplex between them in addition to the 1-simplices 
(the edges) connecting them. Of course, the model we develop here could be generalized 
very naturally beyond 2-simplices but we postpone this more involved generalization to 
future work.\medskip

In this work, we define the simplicial adaptive voter model and study
its dynamics numerically by direct simulation. On the one hand, our model turns out to 
preserve some key features of the standard adaptive voter model regarding metastability
and diffusive absorption~\cite{DemirelVazquezBoehmeGross} into a single opinion for low re-wiring as well as a
fragmentation transition for high re-wiring. The quantitative structure of these
transitions turns out to be significantly affected by peer-pressure. In particular,
absorption occurs faster, and the fragmentation transition occurs earlier 
with respect to the re-wiring frequency. From the viewpoint of opinion dynamics,
this can be interpreted in the sense that societies are driven faster into
mono-opinion/polarized or fragmented opinion states if peer-pressure effects occur. These 
effects could evidently be induced from social network interactions, i.e., there
is a potential danger that we are going to observe fragmented or polarized
societies much earlier/faster than classical models would anticipate. Furthermore,
we also find a highly interesting mathematical effect in the simplicial adaptive
voter model. It does happen frequently that due to re-wiring of edges, even with 
replacement of lost simplices, that there are eventually no active simplices with
different opinions left. This effect tends to occur before the final asymptotic 
dynamics of the voter process has been reached, i.e., when there are no active edges
left with different opinions. This leads to the conjecture that dynamical models
on simplicial complexes can display a multiscale~\cite{KuehnNetworks1} hierarchy, where 
higher-order simplices equilibrate before lower-order simplices 
do~\cite{DemirelVazquezBoehmeGross,KuehnMC}.

\section{The Adaptive Simplex Voter Model}
\label{sec:asvm}

In this section, we introduce the variants of the voter models in more detail. We 
are going to provide some basic background and references and then define the 
simplicial adaptive voter model.\medskip 
 
Consider a simplicial 2-complex $\cS$, which consists of zero-dimensional 0-simplices 
(or vertices) $\cV$, one-dimensional 1-simplices (or edges) $\cE$ and two-dimensional 
2-simplices $\cT$. Recall that for a simplicial complex one requires 
that each face of a simplex is again in the simplicial complex, and that the non-empty
intersection of two simplices is a face for each of the two simplices. In our modelling 
context this makes particular sense since a triadic friendship does generally also
contain friendships between the respective three individuals. These friendships are represented 
by the faces of the 2-simplex, which are the edges. We want to define a minimal adaptive 
voter model on the space of simplicial complexes with vertices labeled by two possible 
opinions. For notational simplicity and convenience we allow for two states $-1$ and $1$
and use the labels
\benn
\X=\text{ vertex of state $-1$},\quad \text{and}\quad \Y=\text{ vertex of state $1$.}
\eenn 
The possible edges 
\benn
\YY\quad \XY\quad \XX
\eenn
are either state-homogeneous (\textit{inactive}) or state-inhomogeneous (\textit{active}). 
The 2-simplices can occur in any of the following four configurations
\benn
\YYY\quad \YYX\quad \YXX\quad \XXX
\eenn 
where the first and the last one are state- and edge-homogeneous, while the second and 
third are state- and edge-heterogeneous. Note that we use double-edges to indicate a 
2-simplex in comparison to a triangle, i.e., a full subgraph on three 
nodes, which is not part of a 2-simplex. The interior of a 2-simplex is color-coded with 
current majority opinion within the triangle.\medskip  

Let us recall the classical adaptive voter model on a graph $(\cV,\cE)$. At each time step 
an edge $e\in\cE$ is chosen at random and one of the following possibilities can then occur:

\begin{itemize}
 \item[(C1a)] (``social avoidance''): If $e$ is active, then with probability $p\in[0,1]$ 
one of the vertices (chosen with equal probability) re-wires the edge to a vertex with its 
own opinion, which is chosen at random from the remaining vertices. We can represent this
rule graphically by: 
\benn
\XY \X\mapsto \Y\XX\qquad \YX \Y\mapsto \X\YY.
\eenn  
Note that the probability $p$ is a very crucial parameter in the adaptive voter model.
 \item[(C1b)] (``personal discussion''): If $e$ is active, then with probability $1-p$ 
one of the two vertices of the edge $e$ is chosen at random with probability $1/2$ and it 
adapts the opinion of the other vertex:
\benn
\XY \mapsto \XX\qquad \XY\mapsto \YY.
\eenn 
\item[(C2)] (``inert situation''): If $e$ is inactive, then nothing happens.
\end{itemize}

Of course, the rules imply a conservation law of the number of edges as well as the 
number of vertices, which tends to be helpful to reduce dimensionality, simplify the
mathematical analysis, and to benchmark computations by checking whether the conservation
laws hold. Next, we describe a minimal model extension to include the role of $2$-simplices, 
which is also meaningful for applications. Consider the simplical complex $(\cV,\cE,\cT)$ 
and another probability $q\in[0,1]$, that encodes the peer pressure. Again we select an 
edge $e$ at random at each event time step. Then we use the following rules:
 
\begin{itemize}
  \item[(R0)] If $e$ is not part of a 2-simplex, then the classical rules apply. 
  \item[(R1)] If $e$ is part of at least one 2-simplex and of type $\XY$ then with 
	probability $q$ a majority rule is implemented and with probability $(1-q)$ the classical 
	rule (C1) is implemented. When the majority rule applies, one of the simplices attached 
	to $e$ is chosen and the majority persuades the minority with probability $p$
  \benn
  \XXY \mapsto \XXX \qquad \XYY \to \YYY.
  \eenn
  When the rewiring is chosen, irrespective of the peer-pressure, all simplices attached 
	to that edge break up, e.g., 
  \benn
  \XYX \Y\mapsto \xxyy \qquad \YXY\X\mapsto \yyxx
  \eenn
  and triangles are chosen randomly for conversion into a 2-simplex
  \benn
  \xxx\to\XXX \quad \xxy\to\XXY \quad \xyy\to \XYY \quad \yyy\to \YYY
  \eenn
  to preserve the total number of simplices.
  \item[(R2)] If $e$ is part of at least one 2-simplex and not of 
	type $\XY$ it stays inert.
\end{itemize}

The rules (R0)-(R2) are a very natural extension of the classical adaptive voter
model rules (C1)-(C2). Again, conservation of vertices and edges is guaranteed.
The new rule (R1) tries to conserve simplices as long as possible. However, due 
to re-wiring, one may eventually not have any triangles left that can be
converted into 2-simplices. A subsequent rewiring event within a 2-simplex would 
therefore reduce the overall content of 2-simplices and thus violate the conservation 
of simplices. This leaves two natural options:
\begin{itemize}
 \item We stop the simulation precisely at the first time when the triangles have 
been depleted.    
 \item We continue with the simulation despite the violation of the simplex 
conservation until an absorbing state of the Markov chain is reached.
\end{itemize}

We will always indicate in our simulations, which option we have chosen. Practically,
all our main new results are just focusing on using the first option as it provides 
us with the regime, where the new 2-simplex rules are relevant.

\section{Results}
\label{sec:results}

Before describing our simulation results for (R0)-(R2), we briefly recall the 
well-known results for the classical adaptive voter model (C1)-(C2) on a graph.

\subsection{The Clasical Co-evolving Voter Model}

\begin{figure}
\begin{center}
 \includegraphics[width=\linewidth]{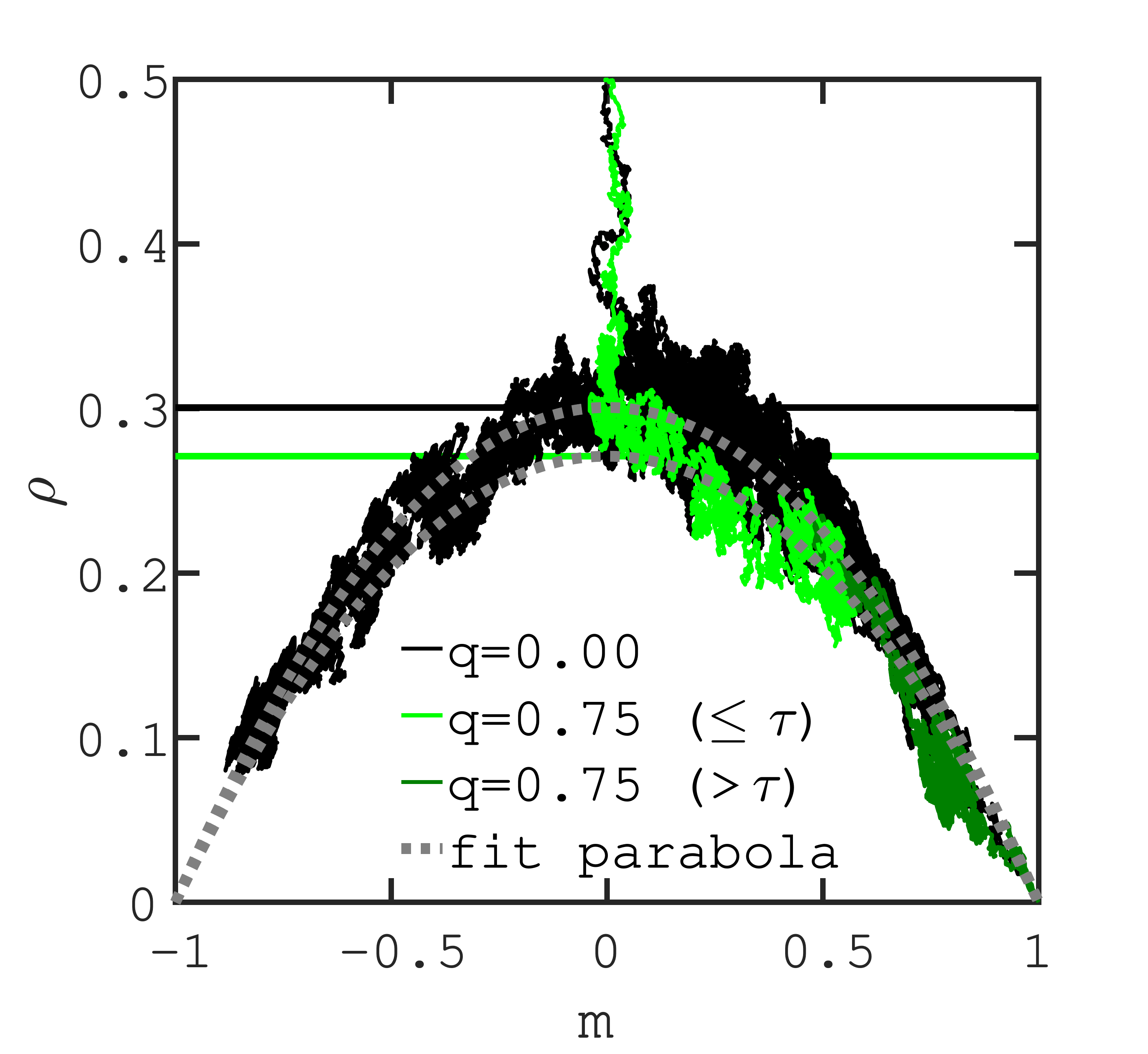}
 \caption{We show two sample paths of the dynamics in the 
$(m,\rho)$-space at a rewiring probability of $p=0.55$: The black path shows the 
scenario where peer-pressure is absent, i.e. $q=0$. The light and dark green paths shows the scenario 
where the peer-pressure is at $q=3/4$, respectively before the depletion of triangles at time $\tau$ and thereafter. The dotten grey lines are best fits of the paths to the parameterized parabola $\rho=\xi_p(1-m^2)$, respectively for $q=0$ and $q=0.75$ before the depletion of triangles. The horizontal lines indicate the respective values of their apexes $\xi_p$. The size of the network is $N=500$, its mean 
degree is $\mu=8$ and the simplex-per-edge degree is $s=0.2$. There is an initial 
population of triplets, such that the triplet-per-edge density is 0.8.}
\label{fg:sample}
\end{center}
\end{figure}

The classical co-evolving voter model \cite{VazquezEguiluzMiguel,DemirelVazquezBoehmeGross} 
corresponds to $q=0$. There one observes two different phases --- the active and the frozen 
phase --- along the parameter $p$, separated by a fragmentation transition at $p_c$. In the 
active phase ($p<p_c$) the dynamics evolves towards a slow manifold of strictly positive 
active edge densities and then follows a random walk along this manifold towards 
a state with a giant component, all of whose members are of the same state and all 
of whose active edges have consequently vanished; cf.~Figure~\ref{fg:sample}. 
In the frozen phase ($p>p_c$) the 
dynamics evolves towards a fragmented state in which two disconnected and internally 
state-uniform components exist. These two phases still exist in the simplical 
co-evolving voter model. This shows that our model is really a minimal extension as
several main effects are preserved. We are interested in the behaviour of the transient 
and limiting behaviour as $q$ is deformed away from zero. When $q>0$ the simplices start 
to have an effect, because the majority rule inside a simplex is governed to some 
extend (i.e., via $q>0$) by the peer pressure.\medskip 

It was observed in the classical case that it is helpful to view the dynamics within a 
compact region under a suitable projection \cite{VazquezEguiluzMiguel}. To this end 
let $\sigma_+$ and $\sigma_-$ denote the 
relative densities of the two opinions $\pm1$ across all nodes. Note that 
$0\leq\sigma_\pm\leq 1$ and $\sigma_++\sigma_-=1$. Then we denote the difference in
opinion, i.e., the majority disparity by $m=\sigma_+-\sigma_-$. In a statistical physics context
we can also draw the analogy of $m$ to the magnetization, e.g., when thinking of the 
classical Ising model. Furthermore, we denote the active link density by $\rho$. It is very
helpful to use the coarse-grained $(m,\rho)$-coordinates to understand the dynamics. The 
network initially looses active links. Either all active links are depleted directly 
without any of the opinions becoming dominant in the process or the active links reach 
a quasi-stationary density at a positive value. There it enters into a random walk on 
a neighbourhood of a parabolic-shaped region defined via the relation $\rho=\xi_{p}(1-m^2)$, 
for some constant $\xi_p$ indexed by $p$. In that case one opinion may gain the majority, 
as $m$ deviates from zero along that region. Which of these two scenarios happens depends 
on whether $p<p_c$ or $p>p_c$; see Figure~\ref{fg:sample}. On the parabola the active edge 
density evolves much slower than initially, which is why one may refer to this region as the 
slow or inertial manifold. Eventually the random walk hits either of the end points 
$(m,\rho)=(\pm 1,0)$, corresponding to a giant component of a single opinion. The 
value $\xi_p$ is a characteristic $p$-dependent value of the slow manifold and corresponds 
to the quasi-stationary density of active links at vanishing majority disparity. In \cite{VazquezEguiluzMiguel} they consider the time-average of all the quasi-stationary densities 
of surviving runs $\rho_{surv}$, which is then taken as the order parameter of the model, 
however $\xi_p$ is another legitimate choice.\medskip

In this work, we perform a numerical study of the $q$-deformed simplical co-evolving voter 
model. In the following we describe the implementation and the results.

\subsection{Initialization}

First we initialize a random simplicial complex $(\cV,\cE,\cS)$ and assign an equal 
amount of $+1$ and $-1$ states to the vertices at random. We want to construct this 
complex from the data consisting of the number of vertices $N$, of edges $E$ and of 
2-simplices $S$, or alternatively the mean degree $\mu=2E/N$ and the 2-simplices-degree 
per edge $s=3S/E$. An important aspect of the dynamics (R0)-(R2) is the 
simplex-preserving transformation of a triangle into a 2-simplex, whenever a 2-simplex 
is destroyed by a rewiring action. Due to these transformations one also requires an 
initial population of triangles in the network, an amount $T$, so that a significant 
fraction of edges should be part of triangles. There are at least 
two ways to create a random simplicial complex from the data $(N,E,S,T)$. One option 
is to pick $E$ edges uniformly at random from the list of $\binom{N}{2}$ possible 
vertex pairs. However, there is a chance that not enough triangles are created to 
declare $S$ of them as 2-simplices and $T$ as triangles. Another method is to pick 
$S+T$ triangles from the $\binom{N}{3}$ combinations of unordered non-repeating 
three-tuples. An amount $S$ are declared 2-simplices and the rest, i.e. $T$, remain 
triangles. All edges that were thus created and are part of triangles or 2-simplices 
become part of the edge set $\cE$. If there are not yet $E$ edges, ones picks uniformly 
at random the remaining edges from vertex pairs that are not yet part of the edge 
set. If more than $E$ edges were introduced already by forming triangles and simplices, 
the method has failed. The first method tries to reduce degree correlations and works 
well for large $\mu$ and very low $s$. The second method aims to reduce correlations 
between the number of simplices per edge at the cost of degree correlations and is 
guaranteed to work when $3(S+T)\leq E$. We choose the second method to allow for 
larger values of $s$. In the Appendix~\ref{ap:algorithm} we show the details of this 
method.

\subsection{Phase Portrait}

\begin{figure}
\begin{center}
 \includegraphics[width=\linewidth]{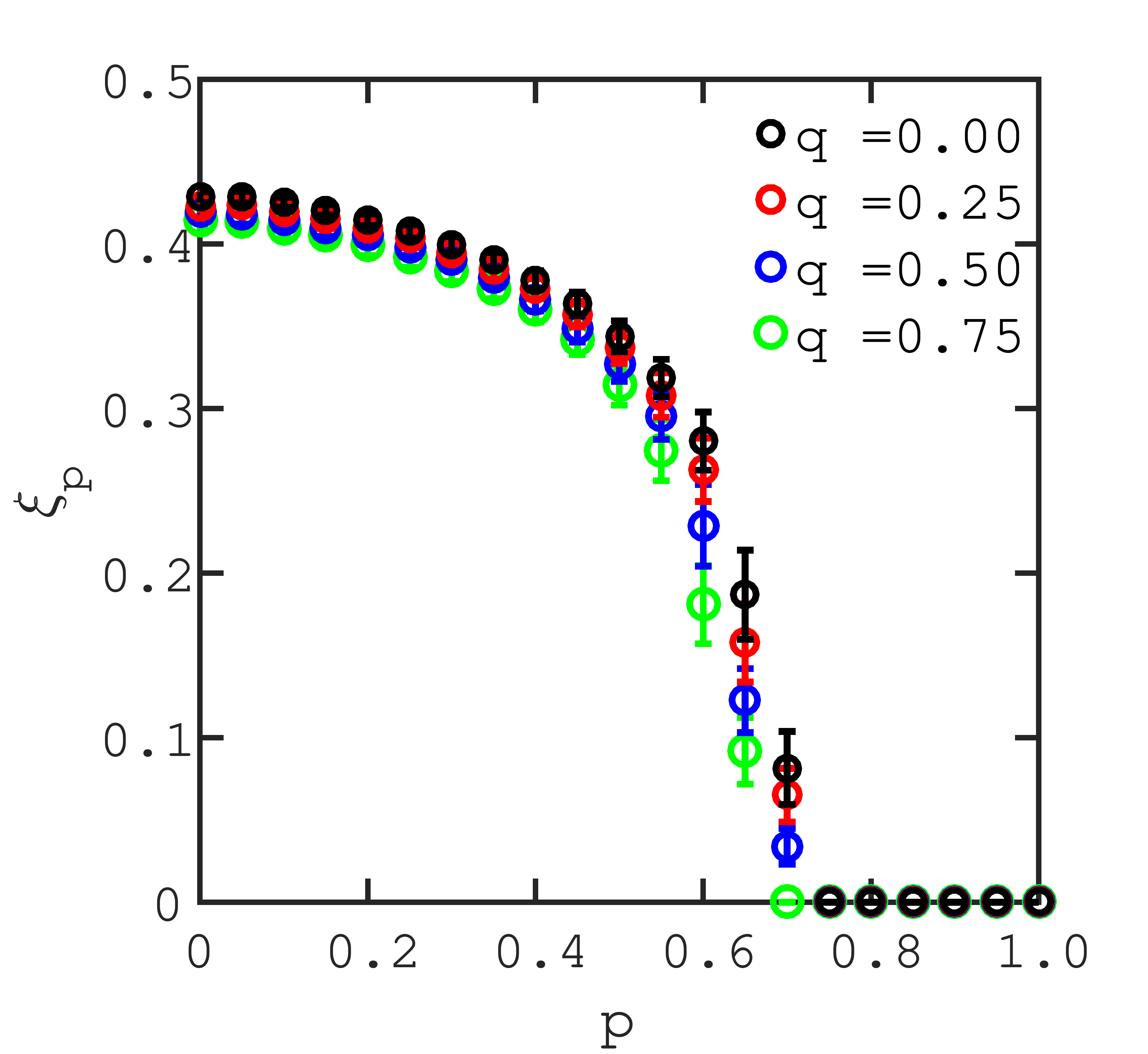}
 \caption{We compare 
the the phase portraits for various peer-pressures. The estimates of the order parameters $\xi_p$, i.e. the apexes of the parabolic regions (c.f. Figure \ref{fg:sample}), are plotted for a range of rewiring probabilities between 0 and 1 and peer-pressures $q\in\{0,0.25,0.5,0.75\}$, which are respectively color-coded with black, red, blue and green. Each data point is generated from 200 runs and shows the mean and the standard deviations of the fitted values for $\xi_p$. Here, the size of the network 
is $N=500$, its mean degree is $\mu=8$ and the simplex-per-edge degree is $s=0.2$. We 
also have an initial population of triplets, such that the triplet-per-edge density 
is 0.8.}
\label{fg:phases}
\end{center}
\end{figure}

For $q>0$ the network dynamics also falls onto a parabola-shaped region in $(m,\rho)$-space, 
on which the active edges evolve much slower. This is exemplified in Figure~\ref{fg:sample}, 
where we compare two sample paths with peer pressures $q=0$ and $q=0.75$ at a rewiring rate $p=0.55$. However, 
the presence of the peer pressure lowers the quasi-stationary densities. This effect 
can be explained by the enhanced force to eradicate active edges: Consider a heterogeneous 
2-simplex in which one node is of opinion $+1$ and two of opinion $-1$. It has two active 
edges and suppose one is chosen for an update. If the node with the minority opinion 
convinces the neighbor with the majority opinion, then there are still two active edges 
in the simplex. If on the other hand the majority convinces the minority node there are 
none. Thus, the higher the probability of a majority rule, the higher the tendency to 
reduce active edges. The same argument holds of course for a 2-simplex with the opposite 
majority. This effect happens irrespective of system size, edge- or 2-simplex-densities, 
as long as they are positive. 

In Figure~\ref{fg:phases} we show the phase portraits for various values of $q$ with 
$N=500$ at an edge-per-vertex degree $\mu=8$ and a simplex-per-edge degree $s=0.2$. 
We observe that the peer pressure shifts the critical threshold to lower 
rewiring probabilities. This also follows from the previous observation. If the active 
edge densities are reduced by the majority rule, then the active edges 
will vanish at lower rewiring probabilities. This means that the slow manifold is never 
reached and consequently a fragmentation takes place rather than a random walk towards 
any of the single-opinioned final states. We also note that the variance of the order parameter increases towards the fragmentation threshold, as is expected. 

\subsection{Depletion of Triangles}
\label{sc:depletion}

\begin{figure}
 \includegraphics[width=\linewidth]{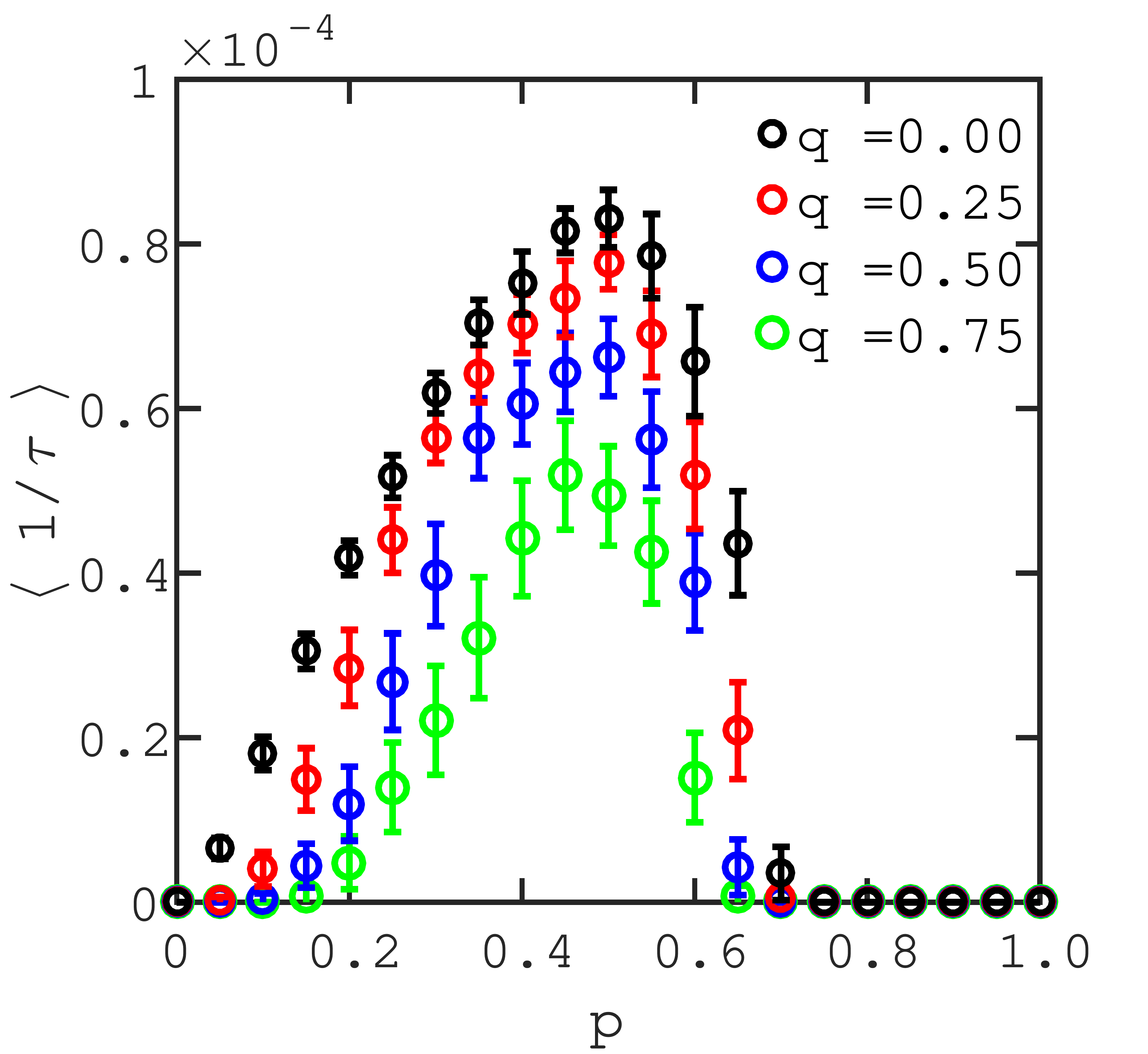}
\caption{We show the average inverse depletion time of triplets for rewiring 
probabilities in the entire range between 0 and 1 and for peer pressures $q\in\{0,0.25,0.5,0.75\}$. All other parameters, i.e. $N,\mu,s$ 
are as above.}
\label{fg:tripdepl}
\end{figure}

In the classical co-evolving voter model there is a depletion of active edges.
One crucial feature of the co-evolving voter model on a simplicial complex is 
the depletion of the higher-order structures, which in our case are the 2-simplices 
and the triangles. Heterogeneous simplices are either homogenized by the peer pressure 
or destroyed via rewiring. As simplices are destroyed new ones are created by converting a 
randomly chosen triangle into a 2-simplex. In some parameter regimes this process depletes triangles 
at a higher rate than their production via rewiring. Thus, in these regimes there is a finite 
first triangle-depletion time $\tau$, at which no triangles are left for conversion into 
2-simplices.

In Figure~\ref{fg:tripdepl} we show the average inverse triangle-depletion time $\langle 
\tau^{-1}\rangle$ for a range of rewiring probabilities and peer pressures. 
A value of zero implies an infinite depletion time, or one that is as long as the duration of a simulation. Of course for $p=0$, in the absence of rewiring events, the depletion time is infinite. 
As the rewiring probability is increased, we observe a rise of $\langle 
\tau^{-1}\rangle$ up to a point where a maximum is reached, succeeded by a drop back to zero around the fragmentation threshold. This qualitative behavior can be observed for all values of $q$, with the additional effect that $\langle 
\tau^{-1}\rangle$ is lowered as $q$ becomes larger. In the following we explain this behaviour. 

First, we note that there is a net-decay of triangles only when on average more of them are destroyed than produced. The only way that a destruction or production of triangles can occur is by way of rewiring events. We also remark that the chance for a rewiring to turn a triplet of vertices with two edges into a triangle with three edges rises as the mean degree of the network increases. If at the same time the rewiring link, prior to its rewiring, has a lower probability of being part of a triangle, then there is a net production of triangles.

With this in mind we can explain, both the rise and the fall of the curves as $p$ is increased. When $p$ is getting close to the fragmentation transition, the density of active edges $\rho$ decreases. Consequently, the density of $++$-links ($--$-links), say $\rho_+$ (resp. $\rho_-$), is higher than in uncorrelated networks and so is the mean degree $\mu_\pm=2\rho_\pm E/N_{\pm}$ of the subgraphs with $+1$ states (resp. $-1$ states), where $N_{\pm}$ are their respective sizes. An approximation of the respective mean degrees $\mu_\pm$ via the simplifying assumption $\rho_+/\rho_-\approx N_+^2/N_-^2$, corresponding to an uncorrelated scattering of edges on the two components, and the identity $Nm=N_+-N_-$ yields $\mu_{\pm}\approx \mu (1-\rho)(1\mp m)$ to first order in $m$. See the appendix \ref{ap:connectivities} for a derivation. This relation supports the intuition that the minority component is becoming more densely connected as active edges are depleting and $|m|$ is increasing. Therefore, as the active link density decreases a rewiring event has an increasing probability to produce a triplet rather than to destroy one, given the sparseness of active edges. This effect is enhanced as $p$ is approaching the fragmentation transition with low quasi-stationary values of $\rho$. There, the production may compete with the loss leading to a positive stationary abundance of triangles and a diverging triangle depletion time $\tau$. This effect is of course more pronounced for higher peer pressures as they tend to reduce $\rho$ even further.

We now argue for the initial rise of $\langle 
\tau^{-1}\rangle$. As long as triangles are not produced to a level that compensates their loss, as it happens close to the transition, there is an exponential depletion of them after a finite time. The rate of this depletion depends on the rewiring probability. The higher $p$, the faster they deplete. This effect is opposing the one mentioned before. One may think of it as follows: The values of $\rho$ and $m$ determine how many triangles plus simplex-declared triangles, i.e. $T+S$, the network can maintain on average. The arrival at that quasi-stationary state happens at an exponential rate that is of course proportional to $p$. If, however, that state can only support less than an amount $S$ of triangles and 2-simplices, then the triangles will die very likely and will do so much quicker when $p$ is higher. The effect of peer pressure is that $\rho$ is decreased and the majority disparity in the network is enhanced. Both of these imply an increased production of triangles and thus a longer triangle depletion time.

Finally we also note that beyond the fragmentation threshold the dynamics hits the absorbing fragmented $\rho=0$ state directly without entering the parabolic quasi-stationary region. Therefore it hits an absorbing state ever quicker as the rewiring probability $p$ takes ever larger values. In that regime one cannot make very meaningful statements about the triplet depletion time $\tau$, since it is bounded by the extinction time of the active edges.

We now discuss some direct consequences of the triangle depletion. The conservation of 
simplices is lost once there are no convertible triangles left. This implies also that the simplex-degree per edge $s$ decreases and in turn results in a weakening net effect 
of the peer pressure up to a point, where possibly all 2-simplices are gone. Thus, $s$ ceases to be a fixed parameter of the model. In order to examine the pure effect of the peer pressure $q$ at a given parameter set $(N,\mu,s)$ we therefore measure quantities only
as long as the number of simplices is conserved, i.e., until the triangle depletion time $\tau$. 

Consider a rewiring rate below the fragmentation transition, i.e., $p<p_c$. If $\tau$ is less 
than the time at which the slow manifold is reached, then one cannot take unbiased data from 
the slow manifold, as 2-simplices have already diminished. When triangles are depleted after 
reaching the slow manifold, but before reaching an absorbing state, then one may take unbiased 
data from the slow manifold for times less than $\tau$. This data is suitable for computing the apex $\xi_p$ of the parabola that is fitted through the sample path in the $(m,\rho)$-plane. It is not suitable for extracting the average quasi-stationary active link density of surviving runs, $\rho_{\text{surv}}$, because systematic biases towards higher values would be introduced when data is only taken until some time $\tau$. Consequently we plot $\xi_{p}$ rather than $\rho_{\text{surv}}$.

\subsection{Depletion of Active edges}

\begin{figure}
 \includegraphics[width=\linewidth]{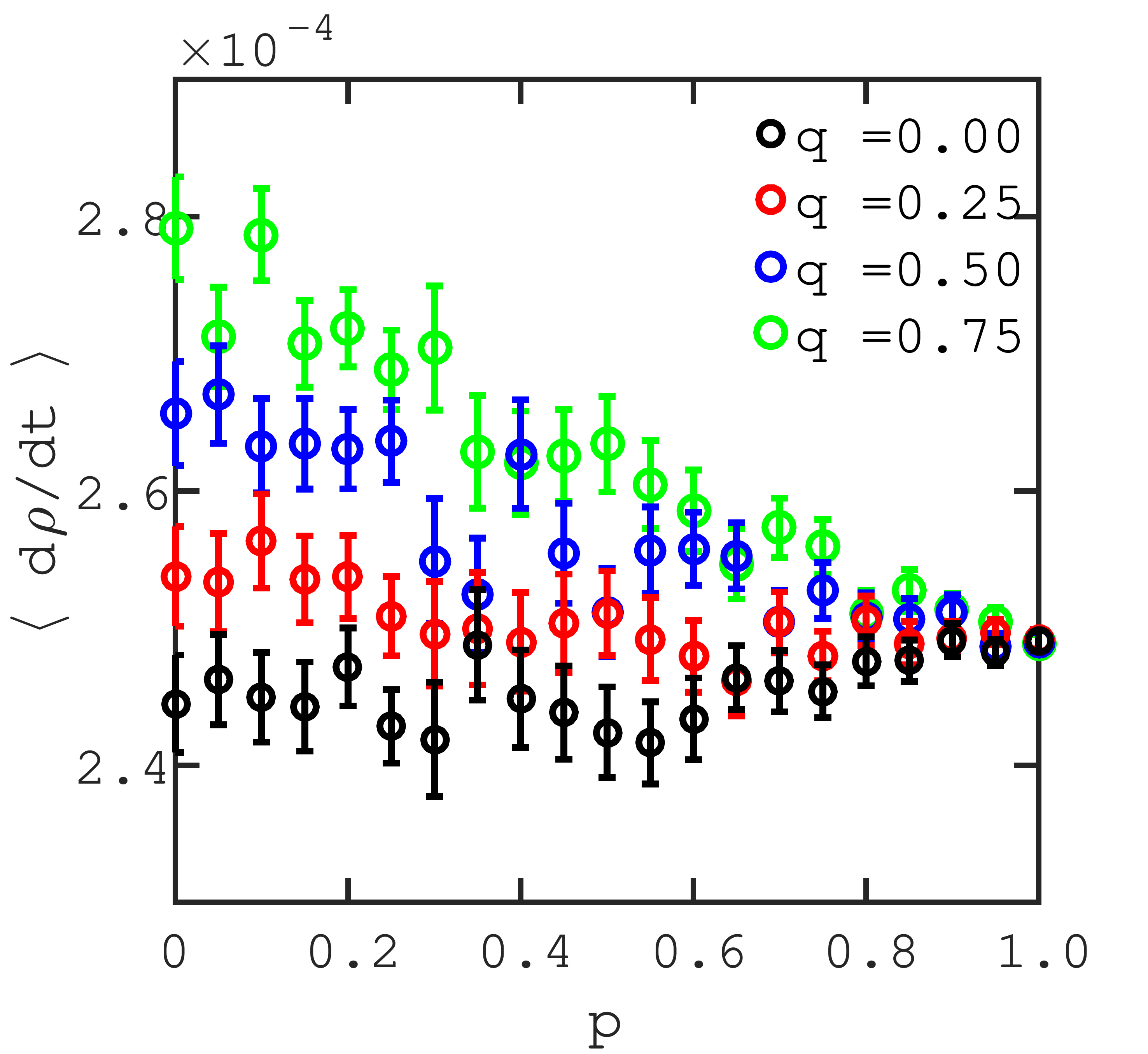}
 \caption{We show the average initial depletion rate of active edges for rewiring probabilities in the entire range between 0 and 1 and for peer pressures $q\in\{0,0.25,0.5,0.75\}$. All other parameters, i.e. $N,\mu,s$ are as above.}
 \label{fg:aldepl}
\end{figure}

There is a fast evolution, before the dynamics reaches either the slow parabolic 
manifold or a fragmented state. During this fast evolution active edges are converted 
into inactive edges by persuasion until the quasi-stationary density is reached. In 
Figure~\ref{fg:aldepl} we plot the average initial depletion rate of active edges for 
various values of $p$ and $q$. Higher peer pressures enhance the depletion rate, 
as expected. This effect is diminished for higher rewiring rates up to the point $p=1$ 
where persuasion does not exist anymore and the peer pressure has no effect.

\subsection{Diffusion and Drift Velocity}

\begin{figure}
 \includegraphics[width=\linewidth]{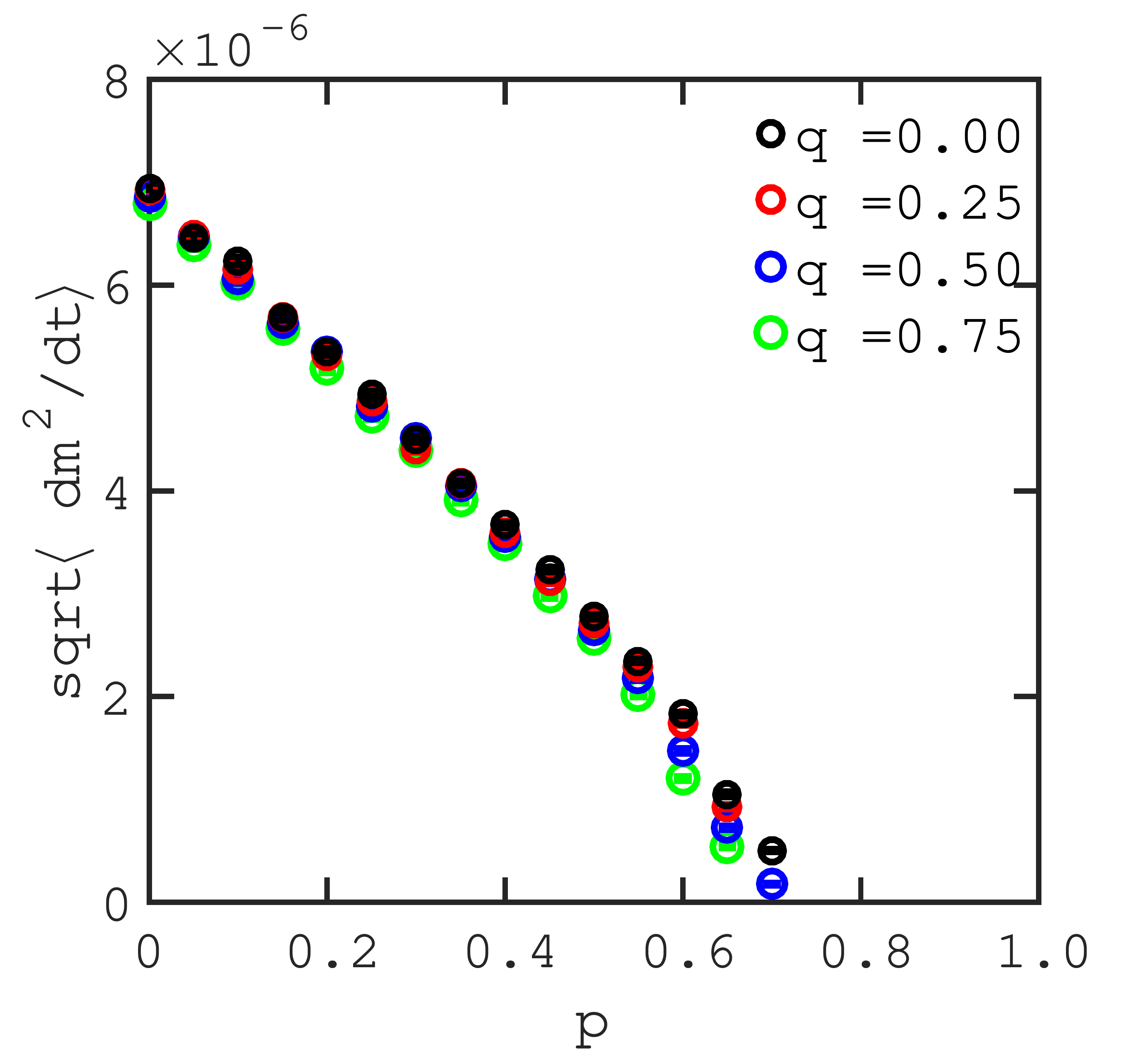}
\caption{This plot shows the diffusivity of the dynamics along the direction of $m$ at $m=0$. It is measured in terms of the mean rate of change of $m^2$ for rewiring probabilities in the range between 0 and 1 and for peer pressures $q\in\{0,0.25,0.5,0.75\}$. All other parameters, i.e. $N,\mu,s$ are as above.}
\label{fg:diffusion}
\end{figure}

We have seen that the peer pressure enhances the depletion rate of active edges during the fast dynamics either towards fragmentation, in the regime $p>p_c$, or before reaching 
the slow manifold, for $p<p_c$. Here, we study
the transient behavior towards the absorbing states. As can be seen in Figure~\ref{fg:sample} the parabolic region is not as well explored for $q=0.75$ as it is for $q=0$. For $q=0.75$ it can be seen often, as in the figure, that only one side of the parabola is visited, indicating that a majority grows once a bias towards one opinion exists. We investigate this behavior by looking at the average rate of change of the majority disparity and its squared value, which are proxies for drift and diffusion respectively. A pure Brownian motion has a vanishing drift. Therefore, we are interested in the extend to which the process is not like a pure Brownian motion. A process with a higher drift is expected to hit the absorbing states at $m=\pm1$ earlier. We do not measure the hitting times directly, however, because they may be larger than $\tau$ (c.f. Subsection \ref{sc:depletion}).

Consider a sample path $(m_t,\rho_t)$\footnote{Formally a sample path $\omega_t$ is a dicrete time-indexed path in the state space $\Omega$, which in this case is the space of $\{-1,1\}$-labelled simplicial complices $(\cV,\cE,\cS)$ and $m$ and $\rho$ are random variables on $\Omega$.}. 
The mean rate of change of the squared distance from $m=0$ at time $s$ is given by 
\begin{equation}
\left\langle\frac{\textnormal{d}}{\textnormal{d} t}\Big|_s m_t^2\right\rangle.\label{eq:diffusion}
\end{equation}
We are interested in this quantity not at any arbitrary point, but precisely as the dynamics is on the slow manifold and has vanishing $m$. Hence, we consider this average, conditional on the event that $m_s=0$ for some time $s$ at which the slow manifold has been reached. For a one-dimensional Brownian motion along the $m$-direction this quantity 
would be twice the diffusion constant $D$. In our simulations we measure it as follows. First we evolve the dynamics until the quasi-stationary parabolic region is reached. After that, whenever the dynamics passes through the $m=0$ line we approximate the average rate of change \eqref{eq:diffusion} via finite differences at that time instance. This is then repeated for many runs and initializations, keeping the parameters fixed.

In Figure \ref{fg:diffusion} we show, how the diffusion 
\eqref{eq:diffusion} changes as the rewiring and peer pressure 
are varied.
The first observation is that the peer pressure does not influence the diffusion 
very much. There is even a slight tendency towards lower diffusions for higher peer pressures. As the 
rewiring probability increases the diffusion decreases linearly for all 
peer pressures. There are two reasons that account for this effect, both of which are based 
on the fact that only persuasion can change $m$. First of all, given a certain amount 
of active edges one expects less changes in $m$ for lower persuasion probabilities, i.e., 
higher rewiring probabilities. Secondly, higher rewiring probabilities decrease the quasi-stationary 
level of active edges, so that persuasions cannot yield as much change in majority disparity. 

We also look at the average drift velocity of the majority disparity. Since the definition of 
the dynamics and hence its probability laws are invariant with respect to the discrete symmetry of 
interchanging the opinions, we expect on average no drift velocity at $m=0$. Therefore, we study the 
average drift velocity at some non-zero majority disparity $m^+\neq 0$:
\begin{equation*}
 \left\langle \frac{\textnormal{d}}{\textnormal{d}t}\Big|_{s} m_t\right\rangle\;,
\end{equation*}
where $s$ is such that $m_s=m^+$. We expect that the drift velocities at $-m^+$ should be minus the drift at $m^+$, in the sense of their probabilitiy laws. This is due to the discrete reflection symmetry. Hence the quantity of interested is the radial drift velocity, pointing away from the origin.

\begin{figure}
 \includegraphics[width=\linewidth]{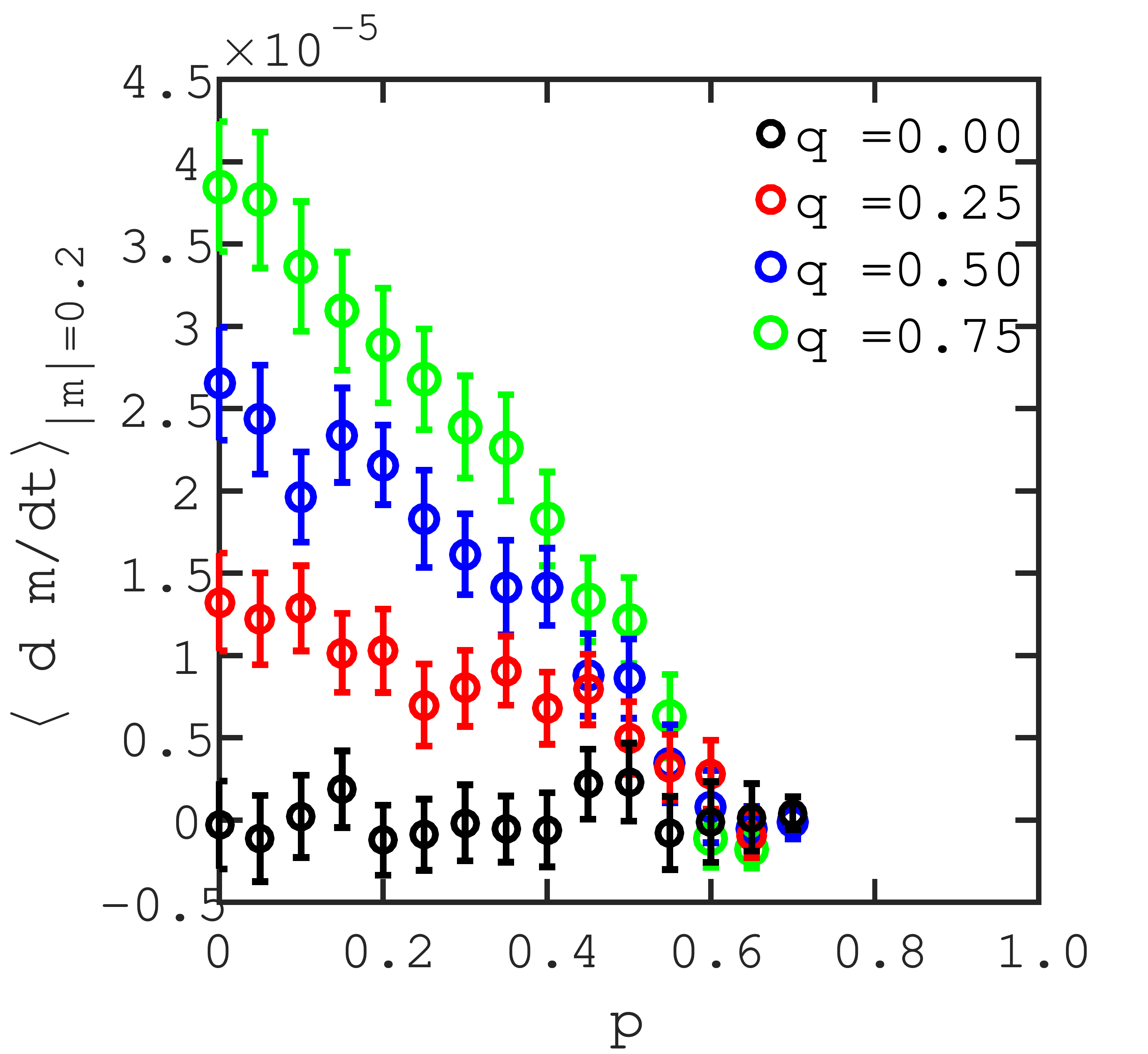}
\caption{We show the average radial drift velocity of $m$ at $m=\pm0.2$ for rewiring probabilities in the range between 0 and 1 and for peer pressures $q\in\{0,0.25,0.5,0.75\}$. There are no data points for rewiring rates above the fragmentation transition, because the dynamics does not enter the slow manifold on which it can explore the regions of higher majority disparity $m$. All other parameters, i.e. $N,\mu,s$ are as above.}
\label{fg:drift}
\end{figure}

In Figure~\ref{fg:drift} we 
plot the average radial drift velocity at $m^+=0.2$. We find that there is no drift velocity at $q=0$. The 
majority disparity in the classical co-evolving voter model is therefore more like a pure diffusion process. 
When $q>0$ we do however observe a strong deviation. The higher the peer pressure $q$ the stronger 
is the average drift velocity towards the extreme points $m=\pm 1$. This effect can be explained by the 
ratio of heterogeneous 2-simplices. When $m_t=m^+$ there are more $\XYY$-simplices than $\XXY$-simplices. 
Thus the majority rule tends to further increase the amount of $\Y$ nodes. The argument holds mutatis 
mutandis for $m_t=-m^+$. These drifts also explain why it becomes more unlikely to change majorities once there is a bias.

In summary, we also see in the slow regime on the parabolic-shaped manifold
a similar effect as in the fast regime. The peer pressure enhances the drift towards
a single majority opinion, i.e., polarization is enhanced. 

\section{Conclusions \& Outlook}

We have shown, how to naturally (from the viewpoint of applications) and minimally (from 
the mathematical perspective) extend the adaptive voter model to a model on
simplicial complexes. It seems now plausible as further steps to also extend other
adaptive contact processes to simplicial complexes, e.g., epidemic spreading models.
Then we demonstrated that the main structural features of the adaptive voter model
remain in the simplicial version, which still yields a fragmentation transition upon
varying the re-wiring rate. Yet, the quantitative properties are changed and we observe
faster transitions to a single-opinion absorbing state or towards a fragmented 
two-opinion state. This is in line with heuristic arguments that peer pressure effects
may lead to polarization; in fact, our model seems to be one of the simplest mathematical
manifestations of this effect. We expect that similar effects do occur in far more 
complicated models involving very large-scale higher-order structures such as interactions
on social networks. Furthermore, we found that the simplicial adaptive voter
model often displays multiple time scales, where the higher-order 2-simplices die out
out before the active edges decay to zero. This multiscale effect leads one to the 
conjecture that simplicial dynamics models could be analyzed order-by-order
with respect to the dimension of the simplices.\medskip

\textbf{Acknowledgments:} LH thanks the Austrian Science (FWF) for support via the grant
Project No.~P29252. CK thanks the VolkswagenStiftung for partial support via a Lichtenberg
Professorship and the EU for partial support within the TiPES project funded the European 
Union’s Horizon 2020 research and innovation programme under grant agreement No 820970.

\appendix 

\section{Algorithm for the Initialization of a Random Simplicial Complex}
\label{ap:algorithm}

Let there be $N$ vertices. We would like to draw uniformly at random $T$ distinct elements from the set of all unordered $K$-tuples without repetition. For the purpose of this paper we are randomly sampling triangles from the vertex list, so that $K=3$, but we expose the method more generally. This set has cardinality $\binom{N}{K}$. Storing such a list is prohibitively large when $N$ is large and $K$ larger than 2 and smaller than $N-1$.  Thus it is advantageous to find an algorithm that needs less storage complexity. 
The idea of the algorithm is to take a natural number $n$ and find the corresponding $K$-tuple in \emph{colex}, which is one of the two canonical orderings defined on unordered tuples without repetitions. We recall that $u=(u_1,u_2,\dots,u_K)$ is less than $v=(v_1,v_2,\dots,v_K)$ in colex\footnote{Unordered $K$-tuples without repetitions are simply sets of the form $u=\{u_1,u_2,\dots,u_K\}$ of cardinality $K$, for which the order of their elements doesn't matter. When $u_i\in\mathbb N$, one may however associate to it uniquely a particular choice of an ordered $K$-tuple $\hat u=(u_1,u_2,\dots, u_K)$ where $u_k<u_\ell$ if and only if $k<\ell$. This is then a unique representation of the unordered tuple $u$ in the realm of ordered tuples, which is of course a much larger space. In abuse of notation we denote the representation $\hat u$ of $u$ in the set of ordered tuples again by $u$. The \emph{colex} order is strictly speaking defined via the representation $\hat u$, rather than $u$.} if and only if $u_k<v_k$ for the last $k$ where $u_k$ and $v_k$ are different.
This mapping from $n\mapsto (u_1,u_2,\dots,u_K)$ is done by iteratively determining the entries, starting from the last one. Suppose the $k+1$-th entry $u_{k+1}$ was found to be $i+1$, where $0\leq i \leq (N-1)$. Then we want to determine the $k$-th entry. To this end we define $F_{i,k}$ as the number of tuples whose $k$-th entry, $u_k$, equals $i$, with all entries above the $k^{th}$ kept fixed, since they are already determined. It is given by $F_{i,k}=\binom{i-1}{k-1}$, because we have to fill $k-1$ slots from a set of $j-1$ numbers. 
The total number of $K$-tuples whose $k+1$-th entry is known to be $u_{k+1}=i+1$ is denoted by $G_{i,k}$. It is simply given by $G_{i,k}=\sum_{\ell=k}^{i}F_{\ell,k}$, which sums up all the tuples whose $k$-th entry is less than or equal to $i$, again keeping higher entries fixed. In other words $G_{i,k}=F_{k,k}+F_{k+1,k}+\dots+F_{i,k}$ and clearly there cannot be any summands where $\ell<k$ in colex. One may find $F$ and $G$ iteratively 
\begin{align}
 F_{i+1,k}&=F_{i,k}\frac{i}{i-k+1}\nonumber\\
 G_{i+1,k}&=G_{i,k}+F_{i,k}\nonumber.
\end{align}
where we use the hockey-stick identity for binomial coefficients.

The last entry, $u_K$, is easy to determine as follows: We note that the first $G_{K,K}$ tuples end with $K$. There is literally only one, namely $(1,2,3,\dots,K)$. The first $G_{K+1,K}$ tuples end with $K$ or $K+1$ and there are $1+K$ of them. The first $G_{j,K}$ tuples end with any number smaller or equal to $j$. Thus we find $u_K$ by searching for $j$ such that $G_{j,K}\geq n > G_{j-1,K}$, or equivalently $\min\{j: G_{j,K}\geq n\}$. Now, that we know $u_K$, our problem reduces to finding the right $K$-tuple in the range between $G_{u_K-1,K}$ and $G_{u_K,K}$. This is equivalent to finding the $K-1$-tuple at position $n^\prime=n-G_{u_K-1,K}$, with the slight modification that these tuples must be taken from the smaller set $\{1,\dots,u_K-1\}$. So now there are $G_{j,K-1}$ such tuples whose $(K-1)^{th}$ entry is $j$ or less. We determine the $(K-1)^{th}$ entry again by finding the minimal $j$ such that $G_{j,K}\geq n^\prime$. We repeat this procedure until all entries have been determined. So in general we have the following iterative algorithm that uses an auxiliary set of variables $\boldsymbol{n}=(n_1,n_2,\dots,n_K)$, initialized as $n_K=n$:
\begin{align}
u_{k}&=\min\{j: n_k\leq G_{j,k}\}\nonumber\\
n_{k-1}&=n_k-G_{u_k-1,k}\nonumber.
\end{align}
We also define $G_{k-1,k}:=0$ to resolve the problem that arises when $u_k=k$. 

\section{Approximation of the subgraph-connectivities $\mu_{\pm}$ in the $(m,\rho)-$space}
\label{ap:connectivities}

Let $\cG$ be a graph with $N$ vertices, $E$ edges and mean degree $\mu=2E/N$. Its nodes are either in the $+1$ or $-1$ states. Let $\cG_\pm\subseteq \cG$ be the subgraph consisting of all the nodes in the $\pm1$ states and their links. We denote the connectivity of $\cG_\pm$ by $\mu_{\pm}$. 
We also define three types of links $(++)$,$(--)$ and $(+-)$ and their respective densities $\rho_+$,$\rho_-$ and $\rho$ as a fraction of $E$. Thus we have
\begin{equation}
\rho_++\rho_-+\rho=1\label{eq:conservation}\;.
\end{equation}
The number of nodes in state +1 or -1 are $N_+$ or $N_-$ and their respective fractions of the entire set of vertices is denoted by $\sigma_+=N_+/N$ or $\sigma_-=N_-/N$. Their difference is denoted by $m=\sigma_+-\sigma_-$. Therefore $N_\pm=\frac{N}{2}(1\pm m)$, which can be used to obtain the following expression for the mean degree in the subgraphs $\cG_\pm$ with $E_{\pm}=\rho_\pm E$ edges:
\begin{equation}
 \mu_{\pm}:=2\rho_\pm E/N_\pm=\frac{4E}{N(1\pm m)}\rho_{\pm}=\frac{2\mu}{1\pm m}\rho_{\pm}\label{eq:mupm}
\end{equation}
So far no assumptions were made about the graph. We would like to approximate the link densities in terms of the effective coordinates $m$ and $\rho$. When $m=0$ a reasonable assumption is $\rho_+=\rho_-$ due to symmetry, which implies by \eqref{eq:conservation} that $\rho_\pm=\frac{1}{2}(1-\rho)$ and consequently that
\begin{equation}
\mu_{\pm}\big|_{m=0}=\mu (1-\rho)\label{eq:zerocase}
\end{equation}
at $m=0$. One may obtain an approximation for the case when $m$ deviates slightly from 0 by requiring that it satisfies 
\eqref{eq:zerocase} for $m\to0$. A first approximation is obtained by making a de-correlation assumption. Suppose we throw $L$ links onto the vertex clouds of $\cG_\pm$, whose total amount of vertex pairs are $P_\pm=\frac{1}{2}N_\pm(N_\pm-1)$ respectively. Thus, on average there will be a total of $L \rho_\pm=L\,P_\pm/(P_-+P_+)$ links in the respective subgraphs and their expected ratio becomes 
\begin{equation}
\frac{\rho_+}{\rho_-}=\frac{N_+(N_+-1)}{ N_-(N_--1)}\approx\frac{N_+^2}{N_-^2}
\end{equation}
Consequently we can plug this approximation into \eqref{eq:conservation} and then \eqref{eq:mupm} and expand the resulting expression to first order in $m$
\begin{equation}
\mu_\pm\approx\frac{2\mu}{1\pm m}\frac{1-\rho}{1+\frac{(1\mp m)^2}{(1\pm m)^2}}=\mu(1-\rho)(1\mp m)+\mathcal O(m^2)\nonumber\;,
\end{equation}
which also satisfies $\lim_{m\to0}\mu_\pm=\mu(1-\rho)$ and resembles \eqref{eq:zerocase}.
Thus for small $m\ll 1$ we can say that the minority component is getting more densely connected, the higher $m$ deviates from 0 and the closer $\rho$ gets to 0. When one of the few active edges rewires in this regime, it will be rewired to one of the two subgraphs $\cG_\pm$ whose higher link densities enhance the chance for triangle production and reduce the chance of triangle destruction due to the sparseness of active links.

\end{document}